\documentclass{article}

\usepackage[nonatbib,preprint]{neurips_2024}

\usepackage[utf8]{inputenc} % allow utf-8 input
\usepackage[T1]{fontenc}    % use 8-bit T1 fonts
\usepackage{hyperref}       % hyperlinks
\usepackage{url}            % simple URL typesetting
\usepackage{booktabs}       % professional-quality tables
\usepackage{amsfonts}       % blackboard math symbols
\usepackage{nicefrac}       % compact symbols for 1/2, etc.
\usepackage{microtype}      % microtypography
\usepackage{xcolor}         % colors

\usepackage{amsmath,amsthm}

\usepackage{hyperref}
    \hypersetup{ colorlinks=true, linkcolor=blue, filecolor=magenta, urlcolor=blue, citecolor=red}

\usepackage[nameinlink,capitalize]{cleveref}

\crefformat{equation}{(#2#1#3)}
\crefname{lemma}{lemma}{lemmas}
\crefname{claim}{claim}{claims}
\crefname{theorem}{theorem}{theorems}
\crefname{proposition}{proposition}{propositions}
\crefname{corollary}{corollary}{corollaries}
\crefname{claim}{claim}{claims}
\crefname{remark}{remark}{remarks}
\crefname{definition}{definition}{definitions}
\crefname{fact}{fact}{facts}
\crefname{question}{question}{questions}
\crefname{condition}{condition}{conditions}
\crefname{algorithm}{algorithm}{algorithms}
\crefname{assumption}{assumption}{assumptions}
\crefname{notation}{notation}{notation}
\usepackage{thm-restate}
  
  \newtheorem{theorem}{Theorem}

  \newtheorem{remark}[theorem]{Remark}
  \newtheorem{corollary}[theorem]{Corollary}
  \newtheorem{definition}[theorem]{Definition}

\usepackage{algorithm}
\usepackage[noend]{algpseudocode}
\usepackage{graphicx}
\usepackage{epstopdf}
\usepackage{subcaption}

\providecommand{\nnreals}{\mathbb{R}_{\geq 0}}

\newcommand{\opt}{\textsf{OPT}}

\def\R{\mathbb{R}}

\def\cO{\mathcal{O}}
\def\cP{\mathcal{P}}

\def\final{0}  % set this to 1 to get a comment-free version
\def\iflong{\iffalse}
\ifnum\final=0  %namely if we allow comments in the output
\newcommand{\rnote}[1]{{\color{red}[{Raymond: \bf #1}]\marginpar{\color{red}*}}}
\newcommand{\enote}[1]{{\color{green}[{\small Elena: \bf #1}]\marginpar{\color{red}*}}}
\newcommand{\ynote}[1]{{\color{purple}[{Young-San: \bf #1}]\marginpar{\color{red}*}}}
\else % in this case [final=1] we don't want any comments to show
\newcommand{\rnote}[1]{}
\newcommand{\enote}[1]{}
\newcommand{\ynote}[1]{}
\fi

\title{A Simple Learning-Augmented Algorithm for Online Packing with Concave Objectives}

\author{%
  Elena Grigorescu\thanks{Supported in part by NSF award CCF-2228814.} \\
  Purdue University\\
  \texttt{elena-g@purdue.edu} \\
  \And
  Young-San Lin \\
  Melbourne Business School \\
  \texttt{y.lin@mbs.edu} \\
  \AND
  Maoyuan Song\thanks{Supported in part by NSF award CCF-2228814, and NSF award CCF-2127806.} \\
  Purdue University \\
  \texttt{maoyuanrs@gmail.com} \\
}

\begin{document}

\maketitle

\begin{abstract}
In recent years, the model of \emph{learning-augmented algorithms} has been extensively studied in the computer-science community, due to the potential of using machine learning predictions in order to improve the performance of algorithms. Predictions may be especially useful for online algorithms, where there is an inherent difficulty in making irrevocable decisions without knowledge of the future. Such learning-augmented algorithms aim to overcome the limitations of  classical online algorithms, in the case when the predictions are accurate, and still perform comparably to the classical online algorithms, when the predictions are inaccurate. 
A common approach is to adapt existing online algorithms to the particular advice notion employed, which often involves understanding previous sophisticated algorithms and their analyses. However, ideally, one would simply use previous online solutions in a black-box fashion, without  much loss in the approximation guarantees.  Such clean solutions that avoid opening up black-boxes are often rare, and may be even missed the first time around.  For example, Grigorescu, Lin, Silwal, Song, Zhou  (\cite{grigorescu2022learning}, NeurIPS 22) proposed a learning-augmented algorithms for the unifying framework of online {\em covering} linear programs, based on the primal-dual method of Buchbinder and Naor (\cite{buchbinder2009online}, Mathematics of
Operations Research, 34, 2009), and on the set-cover approach of Bamas, Maggiori, and Svensson (\cite{bamas2020primal}, NeurIPS 2020). It later turned out that these results can be subsumed by a natural approach  that switches between the advice and an online algorithm given as a black-box, as explicitly noted in \cite{grigorescu2022arxiv}.
This raises the question of whether such simple algorithms may unify other wide families of online algorithms augmented by advice. In this work, we introduce and analyze a simple learning-augmented algorithm for online {\em packing} problems with linear constraints and concave objectives. We believe that this algorithm may be viewed as a general framework for many natural online packing problems. 
In particular, we exhibit several direct applications including online packing linear programming, online knapsack, online resource management benefit, online throughput maximization, and online network utility maximization. Our choice of problem formulation and advice can effectively model those of several prior works.
We further raise the problem of understanding general necessary and sufficient conditions for when such simple black-box solutions may be optimal. We believe this is an important direction of research that would unify many ad-hoc approaches from the literature.

\end{abstract}

\section{Introduction}\label{sec:intro}

In the classical online model, input elements to a problem instance are revealed gradually over time, and an online algorithm is tasked to make irrevocable decisions about the arriving element based on only inputs revealed to it, without any knowledge of the future.
We refer the readers to~~\cite{hoi2021online,buchbinder2009survey} for surveys on the topic.
Performances of online algorithms are often measured in their \emph{competitive ratio}, which is the ratio between the `cost' or `value' of the solution obtained by an online algorithm, and that of the optimal offline algorithm with access to full knowledge of the problem instance in advance. 
Due to the adversarial nature of worst-case analysis, as well as the inability to predict the future, online problems often exhibit impossibility results, namely lower bounds on the competitive ratio of any online algorithm (e.g.~\cite{alon2009online,balseiro2019learning,karlin1994competitive}). 

Recent advances in machine learning have inspired the community to go `beyond the worst-case', utilizing the predictive ability of machine learning models to augment the online algorithm with information about the future based on historical data. Unfortunately, the lack of theoretical guarantees of these models imply that the provided predictions may be arbitrarily inaccurate, and blindly following these predictions can lead to embarrassingly inefficient or invalid solutions, as shown in~\cite{szegedy2013intriguing,bamas2020primal}. As a result, the focus of the community has shifted to devising \emph{learning-augmented algorithms}~\cite{lykouris2021competitive,purohit2018improving} that use such possibly erroneous advice more prudently, and are able to achieve  improved performance when the advice is accurate, a notion called \emph{consistency}, while maintaining reasonable worst-case guarantees when it is not, a notion called \emph{robustness}. We refer the reader to~\cite{roughgarden2021beyond} for a survey in beyond worst-case analysis, and~\cite{mitzenmacher2020algorithms} for a survey in learning-augmented algorithms.

\paragraph{Online covering problems.} One such line of work studies learning-augmented algorithms in the context of online covering problems with linear programming (LP) formulations. Bamas, Maggiori, and Svensson~\cite{bamas2020primal} devised a learning-augmented algorithm for online set cover, using the primal-dual framework of Buchbinder and Naor~\cite{buchbinder2009online}. By adapting these algorithms and the results of Elad, Kale, and Naor \cite{elad2016online}, 
Grigorescu, Lin, Silwal, Song and Zhou~\cite{grigorescu2022learning} solve general online covering LPs and semidefinite programs (SDPs). Their techniques involve solving the covering LP and the dual packing LP simultaneously,  while fine-tuning the growth rate of each covering variable carefully with the use of the advice provided to the algorithm. 

\paragraph{The switching approach.} It was noted in \cite{grigorescu2022arxiv}(Appendix B)\footnote{Attributed to subsequent private communication with Roie Levin.}, however, that there exists a simple learning-augmented algorithm that outperforms their arguably sophisticated primal-dual algorithm, 
for all parameter choices. The algorithm involves a  switching strategy, where it starts by following the solution provided by any state-of-the-art online algorithm, and then switches to the solution implied by the advice, once the cost of the online solution surpasses that of the advice. It is remarkably easy to implement and  analyze. In addition, it uses the subroutine online algorithm as a black-box, absolving the algorithm's designer of the need to understand the possibly sophisticated methodology employed by prior literature. 

\paragraph{When is switching optimal?} Designing simple  solutions to natural problems of wide general interest is an ultimate goal of both theory (as they can be taught even in undergraduate courses!) and practice (as they can be implemented with a few lines of code, and would have strong provable guarantees!). However, while the study of learning-augmented algorithm is currently   flourishing
(see~\cite{ALPS} for an archive of relevant works),
many solutions and formulations are often somewhat ad-hoc, and hence there is a need for general frameworks, techniques, and paradigms. Here we raise a basic question:
\begin{center}
    {\em Can one characterize the space of online problems augmented with advice, for which one may use classical online algorithms as a {\em black-box} to obtain {\em optimal} solutions?}
\end{center}
In particular, what are the necessary and sufficient conditions for the simple switching strategy above to be close to being optimal? What features should a problem exhibit that would allow it to be solvable in a black-box fashion in the advice setting? We believe that a better understanding of this conceptual direction may lead to a unifying framework in the study of algorithms with advice.

In this work, we make progress on sufficient conditions for the question above. In particular, we study a general class of online \emph{packing} problems with advice and show that the switching framework is close to being optimal. 

\paragraph{Packing problems.} Packing problems are a class of maximization problems dual to the general online covering problems. Concretely, we consider problems of the following form:
\begin{align}
  \begin{aligned}
    \text{maximize } & f(x) 
    \text{ over } x \in \nnreals^n 
    \text{ subject to } A x \leq b.
  \end{aligned} \label{opt:packing}
\end{align}
Here, $A := \{a_{ij}\}_{i \in [m], j \in [n]} \in \R_{\geq 0}^{m \times n}$, $f: \nnreals^n \to \nnreals$ is a concave and monotone function such that $f(\delta x+(1-\delta)y) \geq \delta f(x)+(1-\delta)f(y)$ for all $x,y \in \nnreals^n$ and $\delta \in [0,1]$, $f(x) \ge f(x')$ for any $x \ge x'$, and $f(\mathbf{0}) = 0$, and $b \in \R_{> 0}^m$ denotes an upper bound vector for the linear constraints. We use $i \in [m]$ and $j \in [n]$ to denote the row and column index of $A$, respectively.  

In the online packing problem, the values in $b$ and the number of constraints $m$ are given in advance. In each round $j \in [n]$, a new packing variable $x_j$ is introduced, along with all the associated coefficients $a_{ij}$ for all $i \in [m]$ (the $j$-th column of $A$). The number of packing variables, $n$ might be unknown, and each packing constraint is gradually revealed column-wise to the algorithm. In round $j \in [n]$, the algorithm can only irrevocably assign a value for $x_j$. The goal is to approximately minimize $f(x)$ by assigning values to the variables online and maintaining (approximate) feasibility.

\paragraph{Online decision making with advice.} In the learning-augmented online packing problem, we are also given advice $x' \in \nnreals^n$, in an online fashion. In round $j \in [n]$, the advice $x'_j$ is revealed, suggesting what the magnitude of the arriving online packing variable $x_j$ should be. A learning-augmented online algorithm can incorporate the advice to maximize the objective $f(x)$ by irrevocably assigning the values of $x_j$, potentially outperforming any traditional online algorithm. 

To measure the performance of a learning-augmented algorithm, we compare the value of its proposed solution to that of both the advice and the offline optimal solution. A solution from a learning-augmented algorithm is \emph{$C$-consistent} if it is at least $\frac{1}{C}$ the value of the advice, and is \emph{$R$-robust} if it is at least $\frac{1}{R}$ the value of the optimal offline solution. We consider a relaxation of the packing constraints, allowing a solution to approximately violate some or all of them by some factor $V$, i.e., $A x \leq V \cdot b$. In this case, we say that $x$ is \emph{$V$-feasible}. Note that scaling a $V$-feasible solution by a factor of $\frac{1}{V}$ returns a feasible solution. A learning-augmented algorithm is $(C,V)$-consistent if its solution $x$ is $C$-consistent when $x$ is $V$-feasible. Let $\opt$ denote the optimal objective of an offline problem. The following measures the performance of a learning-augmented online algorithm.
\begin{definition}
    A learning-augmented online algorithm for the packing problem \eqref{opt:packing} which takes $x'$ as the advice is $(C,V_C)$-consistent, $R$-robust, and $V$-feasible if it generates $x$ such that
    \[f(x) \ge \max\left\{\frac{f(x')}{C}, \frac{\opt}{R}\right\}\]
    when $x'$ is $V_C$-feasible, $f(x) \ge \opt/R$ when $x'$ is not $V_C$-feasible, and $x$ is $V$-feasible.
\end{definition}

\paragraph{Our contributions.} 
Our main result is a general framework for devising learning-augmented algorithms for online packing problems. 
We present a simple algorithm utilizing a switching strategy, which uses any state-of-the-art classical online algorithm as a black-box subroutine, and obtains a solution that matches both the value of the advice and the value of the subroutine online algorithm asymptotically:
\begin{restatable}[]{theorem}{Main}\label{thm:Main}
For the packing problem \eqref{opt:packing}, there exists a learning-augmented online algorithm that takes an $\alpha$-competitive $\beta$-feasible online algorithm as a subroutine and an advice $x' \in \nnreals^n$. The algorithm is $(2,\beta)$-consistent, $2\alpha$-robust, and $3\beta/2$-feasible.

\end{restatable}

We formally state our algorithm and its analysis in the proof of \Cref{thm:Main} in \Cref{sec:Switch}.

We remark that our algorithm is a general-purpose framework, and does not rely on any specific classical online algorithm. As hinted above, this result absolves users of our framework from the responsibility of understanding potentially sophisticated online algorithm used as a black-box subroutine. In addition, future advancements in the field of classical online problems would immediately imply advances in their variants augmented by advice. 
Problems that fall under the packing maximization problems with concave objective functions  for which our framework may apply includes network utility maximization~\cite{shi2015large,cao2022online}, compressed sensing~\cite{donoho2006compressed,sun2019deep}, channel transmission~\cite{johansson2005resource}, biological inference~\cite{de2018introduction}, low-rank matrix recovery~\cite{recht2010guaranteed,chi2019nonconvex}, auction mechanisms~\cite{zurel2001efficient,kesselheim2013optimal}, associative Markov networks~\cite{taskar2004learning}, etc.

We explicitly study the application of our algorithm to some well-motivated online packing problems including knapsack~\cite{marchetti1995stochastic}, resource management benefit~\cite{leonardi1995line}, throughput maximization~\cite{buchbinder2006improved}, network utility maximization~\cite{cao2022online}, and optimization with inventory constraints~\cite{lin2019competitive} in \Cref{sec:Application}. 

Our advice model on the packing variables also generalizes many prior models (e.g.,~\cite{im2021online,sun2020competitive} in context of online knapsack problems), since any form of advice that suggests a course of action can be simulated by our linear program formulation by setting the advice packing variable to an appropriate value. The exact values of the advice variables depend on the exact problem and advice formulation: We give some examples of these reductions in \Cref{sec:knapsack} for online knapsack. 

Our empirical evaluations show that our theoretical framework is applicable in practice. For online packing LPs, the switching algorithm guarantees robustness and consistency and outperforms both the online algorithm and the advice when the advice only provides a little information.
 
We stress that while our algorithm and analysis is extremely simple compared to that of prior works on learning-augmented algorithms, to the best of our knowledge, the community at large appears to have overlooked the applicability of such simple, and perhaps folklore, algorithms. Thus, we explicitly bring forth the problem of formalizing and solving the unifying questions raised in the introduction. Solutions to these would be a significant conceptual contribution to the area of online learning-augmented  algorithms.

\paragraph{Additional Prior Works}

\emph{Learning-augmented algorithms} have been extensively studied for problems such as set cover~\cite{bamas2020primal}, frequency estimation~\cite{hsu2019learning}, clustering~\cite{ErgunFSWZ22}, scheduling~\cite{lattanzi2020online}, and facility location~\cite{jiang2021online}.
Towards learning-augmented algorithms for general online packing problems, \cite{thang2021online} followed the learning-augmented primal-dual framework of~\cite{buchbinder2009online}, and presented an algorithmic solution. Their model however generalized the objective function to work with multilinear extensions of monotone functions, which incurred a loss in generality. Their advice model is also integral, while our advice allows for fractional predictions.

Other works have studied problems that can be formulated as online packing problems, such as online knapsack~\cite{im2021online}, online matching~\cite{jin2022online}, online bounded allocation and ad-auction~\cite{enikHo2023primal}, and secretary problem~\cite{dutting2021secretaries}. The problem and advice models between these works varies in many aspects, and does not allow a consistent and unifying theme to be found. We remark that our work focuses on general online concave packing problems, which generalizes all of the aforementioned problems, while our form of advice on the value of the packing variables emcompasses all forms of advice which suggests a course of action.

\paragraph{Organization} 
In \Cref{sec:Switch}, we present our algorithm and its analysis.
In \Cref{sec:Application}, we present and discuss several applications.
In \Cref{sec:Experiments}, we present our experimental evaluations on real and synthetic data sets.

\section{The Switching Algorithm}\label{sec:Switch}

We present a simple, possibly folklore,  learning-augmented online algorithm for online packing. Let $\cO$ be an $\alpha$-competitive $\beta$-feasible online algorithm for the packing problem \eqref{opt:packing}, for some $\alpha, \beta \ge 1$.

The algorithm keeps track of two candidate solutions, one from the advice $x'$, and one from the non-learning-augmented online algorithm $\cO$, and combines them. Let $x_j^{\cO}$ be its solution for the $j$-th packing variable. Whenever a column of $A$ arrives online, the algorithm obtains a value for $x_j$ from both the online algorithm and the advice and sets $x_j$ to half of their sum, $(x_j^{\cO} + x_j')/2$. Once the algorithm finds out that the advice violates any constraint by a factor of $\beta$, it discards the advice for this round and follows the online algorithm only. The algorithm is presented in \Cref{alg:simple}.

\begin{algorithm}[!htb]
\caption{A Simple Learning-Augmented Online Algorithm for the Packing Problem \eqref{opt:packing}} \label{alg:simple}
\renewcommand{\algorithmicrequire}{\textbf{Input:}}
\renewcommand{\algorithmicensure}{\textbf{Output:}}

\algorithmicrequire{ An online packing problem \eqref{opt:packing}, the algorithm $\cO$, and the advice $x'$.}

\algorithmicensure{ The online solution $x$.}
\begin{algorithmic}[1]
\For{$j = 1, 2, ...$} \Comment{each arriving row or constraint}
    \State Update $A$ by adding a new column $j$.
    \State Run $\cO$ for round $j$ and obtain $x_j^{\cO}$.
    \If{$A x' \le \beta b$} \Comment{The advice is approximately feasible} \label{line:alg-simple-case-1}
        \State $x_j \gets \frac{1}{2} (x_j^{\cO} + x_j')$. \label{line:xj1}
    \Else
    \label{line:alg-simple-case-2}
        \State $x_j \gets x_j^{\cO}$. \label{line:xj2}
    \EndIf
\EndFor
\end{algorithmic}
\end{algorithm}

\begin{proof}[Proof of \Cref{thm:Main}]
We use \Cref{alg:simple} with the $\alpha$-competitive and $\beta$-feasible online algorithm for $\cO$. Recall that we wish to prove that \Cref{alg:simple} is $(2,\beta)$-consistent, $2\alpha$-robust, and $3\beta/2$-feasible.

The value of the packing solution found by \Cref{alg:simple} is $f(x) \ge f (x^{\cO}/2 + x'_{trim})$, where $x'_{trim}$ is half of the advice $x'/2$ with all entries discarded by the algorithm set to $0$ instead.

It immediately holds that $f(x) \geq f(x^\cO/2) \ge f(x^\cO)/2 \ge \opt/(2\alpha)$, where the second inequality follows by $f(\mathbf{0})=0$ and the concavity of $f$. If $x'$ violates each constraint by at most a multiplicative factor of $\beta$, no entries of $x'$ will be trimmed, and thus $f(x) \geq f(x'/2) \ge f(x')/2$ holds as well.

We observe that $A x'_{trim} \leq \beta b/2$ and $A x^{\cO} \leq \beta b$, which implies that $A x \leq 3 \beta b/2$.
\end{proof}

\begin{remark}
We note that depending on the problem of interest, $\beta$ might not be a fixed parameter. For online packing LPs, $\beta$ is a conditional number non-decreasing over time, which depends on $A$, so it is not necessarily the case that the advice is only used in earlier rounds. See Section \ref{sed:packing-lp} for a more detailed discussion. If $\beta$ is a fixed parameter, the advice is used only in the earlier rounds until it is not $\beta$-feasible. Besides, one can tweak the constant terms of lines \ref{line:alg-simple-case-1}, \ref{line:xj1}, and \ref{line:xj2} in \Cref{alg:simple} to obtain other $(O(1),O(\beta))$-consistent, $O(\alpha)$-robust, and $O(\beta)$-feasible algorithms.
\end{remark}

\section{Applications}\label{sec:Application}

In this section, we present several direct applications of general interest where \Cref{thm:Main} immediately gives non-trivial results, in some cases even optimal, up to constant factors.

\subsection{Online Packing Linear Programming} \label{sed:packing-lp}

The online packing linear programming (LP) problem is a special case of \eqref{opt:packing} where the objective $f(x)$ is the linear function $\sum_{j=1}^n x_j = \mathbf{1}^T x$. Here, $\mathbf{1}$ denotes a vector of ones.
\begin{align}
  \begin{aligned}
    \text{maximize } &  \mathbf{1}^T x
    \text{ over } x \in \nnreals^n 
    \text{ subject to } A x \leq b.
  \end{aligned} \label{opt:packing-lp}
\end{align}

The seminal work by Buchbinder and Naor \cite{buchbinder2009online} presents the following theorem.

\begin{theorem}[\cite{buchbinder2009online}] \label{thm:BN09}
For any $B > 0$, there exists a $B$-competitive algorithm for online packing LP \eqref{opt:packing-lp} such that for each constraint $i \in [m]$ it holds:
\[ \sum_{j=1}^n a_{ij} x_j = c_i \cdot O \left( \frac{\log m \kappa_i}{B} \right), \]
where $\kappa_i=a_i(\max)/a_i(\min)$, $a_i(\max) = \max_{j=1}^n \{a_{ij}\}$, and $a_i(\min) = \min_{j=1}^n \{a_{ij} | a_{ij} > 0\}$.
\end{theorem}
For impossibility results, it is further shown in \cite{buchbinder2009online} that for any $B > 0$, there exists a single-constraint instance for the online packing LP problem, such that any $B$-competitive algorithm must violate the constraint by a factor of $\omega(\log m \kappa / B)$.

The following corollary holds by Theorems \ref{thm:Main} and \ref{thm:BN09}. Here, $\kappa:=\max_{i=1}^m\{\kappa_i\}$.

\begin{corollary} \label{cor:packing-lp}
    There exists a $(2,O(\log m \kappa/B))$-consistent, $2B$-robust, and $O(\log m \kappa/B)$-feasible learning-augmented online algorithm for online packing LP \eqref{opt:packing-lp} that takes parameter $B>0$.
\end{corollary}
\Cref{cor:packing-lp} is optimal in consistency, robustness, and feasibility, up to constant factors. We note that the parameter $\kappa$ is non-decreasing over time. It is possible that Algorithm \ref{alg:simple} enters line \ref{line:alg-simple-case-2} in earlier rounds and then enters line \ref{line:alg-simple-case-1} in later rounds when $\beta=\Theta(\log m \kappa/B)$ increases.

Below we present other specific applications for learning-augmented online packing LP.

\subsubsection{Online Knapsack} \label{sec:knapsack}

In this section, we apply Algorithm~\ref{alg:simple} to the classical online {\em knapsack} problem. In the most basic formulation of this problem, a sequence of $n$ items arrive online, and each item $j$ has value $v_j$ and weight $w_j$. Our goal is to pack a selected set of items with as much value as possible into a knapsack of capacity $C$, deciding irrevocably whether to include each item as soon as it arrives online, without the total weight exceeding the capacity. Formulated as an online packing linear program, we obtain
\begin{equation}
    \text{maximize } \mathbf{1}^T x 
    \text{ over } x \in \nnreals^n 
    \text{ subject to } \sum_{j=1}^n \frac{w_j x_j}{v_j} \leq C. \label{equation:knapsack}
\end{equation}
in which the elements $x_j$ together with it's value $v_j$ and weight $w_j$ arrive online.

As a corollary to \Cref{thm:Main}, it follows that our simple algorithm can be applied to the knapsack problem to obtain asymptotically optimal performance in both consistency and robustness.

\begin{corollary}
Given any $\alpha$-competitive algorithm $\mathcal{O}$ for the online knapsack problem, Algorithm~\ref{alg:simple} using $\mathcal{O}$ as a subroutine implies a $(2,1)$-consistent, $2\alpha$-robust, and $3/2$-feasible learning-augmented online algorithm.
\end{corollary}

The knapsack problem is regarded as one of the most fundamental packing problems in computer science and operations research along with its many variants. It has been extensively studied in the offline setting~\cite{kellerer2004multidimensional,martello1987algorithms} (see~\cite{salkin1975knapsack} for a survey)  
and the online setting~\cite{marchetti1995stochastic, zhou2008budget, ma2019competitive}. Recent work in the learning-augmented setting includes~\cite{sun2020competitive,daneshvaramoli2023online,im2021online,boyar2022online},  which are studying the online knapsack problems with learning augmentation from many angles, such as frequency predictions, threshold predictions, and unit profit settings. 

We point out that our advice model on the values of the arriving dual variables generalizes most, if not all,  forms of advice studied in prior literature. For any advice suggesting a course of action, such as taking an arriving item into the knapsack (or doing so with some probability, in the case of randomized algorithms), we can process such advice by setting the corresponding entry in $y'$ to its value (resp. some fraction of its value).

Another line of work focuses on the advice complexity of online knapsack problems~\cite{bockenhauer2014online,bockenhauer2012advice}, analyzing in fine-grained fashions the amount of information needed about the future online inputs required to devise algorithm that achieves optimality or certain competitive ratios. We point out that advice complexity concerns \emph{trusted} and \emph{accurate} predictions, while the study on learning-augmented algorithms focuses on untrusted, possibly erroneous advice. Thus, the literature on advice complexity is not directly comparable with ours.

\subsubsection{Online Resource Management Benefit}

In this section, we apply Algorithm \ref{alg:simple} to the online resource management benefit problem \cite{leonardi1995line}. For this problem, we have $n$ distinct resources. Each resource $i \in [m]$ can be assigned a maximum capacity $c_i$. Here, $m$ and each $c_i$ are given in advance. A sequence of $n$ jobs is presented online, one job at a time, where $n$ can be unknown. Job $j \in [n]$ has a benefit of $w_j$ and can be scheduled with $r_j$ different alternatives. We use $a^k_{ij}$ for $k \in [r_j]$ to denote the amount of resource $i$ necessary to schedule job $j$ with alternative $k$. Upon the arrival of job $j$, $w_j$, $r_j$, and $a^k_{ij}$ are revealed, and one must schedule job $j$ to different alternatives irrevocably. This problem has the following LP formulation:
\begin{equation} \label{equation:max-benefit}
\begin{aligned}
& \max_{x} & & \sum_{j=1}^n \sum_{k=1}^{r_j} x^k_j \\
& \text{subject to}
& & \sum_{j=1}^n \sum_{k=1}^{r_j} \frac{a^k_{ij} x^k_j}{w_j} \leq c_i & \forall i \in [m],\\
& & & \sum_{k=1}^{r_j} \frac{x^k_j}{w_j} \le 1 & \forall i \in [m],\\
& & & x^k_j \ge 0 & \forall j \in [n], k \in [r_j].
\end{aligned}
\end{equation}
\iffalse
\begin{equation}
\begin{aligned}
& \max_{x} & & \sum_{j=1}^n w_j \sum_{k=1}^{r_j} x^k_j \\
& \text{subject to}
& & \sum_{j=1}^n \sum_{k=1}^{r_j} a^k_{ij} x^k_j \leq c_i & \forall i \in [m],\\
& & & \sum_{k=1}^{r_j} x^k_j \le 1 & \forall j \in [n],\\
& & & x^k_j \ge 0 & \forall j \in [n], k \in [r_j].
\end{aligned}
\end{equation}
\fi
Here, $x^k_j$ denotes the fraction of job $j$ assigned to alternative $k$. Upon the arrival of job $j \in [n]$, the decision $x^k_j$ is irrevocable. 

In \cite{leonardi1995line}, it is assumed that the benefit of each job $w_j \in [1,W]$ with $W \ge 1$ and for every non-zero coefficient $a^k_{ij}$, the ratio $a^k_{ij}/c_i \in [1/P,1]$ with $P \ge 1$. \cite{leonardi1995line} presents an $O(\log mWP)$-competitive algorithm and shows it is the best possible. As a corollary, it follows that with advice, we can achieve the best outcome either obtained from an online algorithm or the advice by losing a constant factor, in a black box manner.

\begin{corollary}
Given any $\alpha$-competitive algorithm $\mathcal{O}$ for the online resource management problem, Algorithm~\ref{alg:simple} using $\mathcal{O}$ as a subroutine implies a $(2,1)$-consistent, $2\alpha$-robust, and $3/2$-feasible learning-augmented online algorithm.
\end{corollary}

\subsubsection{Online Throughput Maximization}

In the Online Throughput Maximization problem \cite{buchbinder2006improved,awerbuch1993throughput}, we are given a directed or undirected graph $G=(V,E)$. Each edge has a capacity $c: E \to \nnreals$. A set of requests $(s_j,t_j) \in V \times V$ arrive online, one at a time. Upon the arrival of request $i$, the algorithm irrevocably chooses (fractionally) an $s_j$-$t_j$ path and allocates a total bandwidth of one unit. The objective is to maximize the throughput. The following packing LP describes this problem.
\begin{equation}
\begin{aligned}
& \max_{x} & & \sum_{j} \sum_{P \in \cP_j}x_{j,P} \\
& \text{subject to}
& & \sum_{P \in \cP_j} x_{j,P} \le 1 & \forall j,\\
& & & \sum_{j} \sum_{P \in \cP_j: e \in P} x_{j,P} \le c(e) & \forall e \in E,\\
& & & x_{j,P} \le 0 & \forall j \quad \forall P \in \cP_j,\\
\end{aligned}
\end{equation}
where $\cP_j$ denotes the set of all the $s_j$-$t_j$ paths. Here, $x_{j,P}$ indicates the amount of flow on the path $P$. The first set of constraints captures that each request requires a unit of bandwidth. The second set of constraints captures that the load on each edge does not exceed the edge capacity.

An $O(1)$-competitive $O(\log |V|)$-feasible online algorithm is presented in \cite{buchbinder2006improved}, which together with our result implies the following.

\begin{corollary}
    There exists a $(2,O(\log |V|))$-consistent, $O(1)$-robust, and $O(\log |V|)$-feasible learning-augmented algorithm for the online throughput maximization problem.
\end{corollary}
\subsection{Online Network Utility Maximization}

In the online network utility maximization (ONUM) problem \cite{cao2022online}, we are given a directed or undirected graph $G=(V,E)$. Each edge has a capacity of one unit. A set of requests $(s_j,t_j) \in V \times V$ arrive online one at a time. Each request $j$ is associated with a budget $b_j \in \R_{>0}$, a monotone concave utility function $g_j: \nnreals \to \nnreals$ with $g(0) = 0$, and a specified $s_j$-$t_j$ path $P_j$. Upon the arrival of request $j$, one must decide irrevocably its allocation rate without exceeding the budget $b_j$. The goal is to approximately maximize the total utility under the online allocation requirement. This problem has the following packing formulation.
\begin{equation}
    \text{maximize } \sum_j g(x_j) 
    \text{ over } x_j \in [0,b_j] \quad \forall j 
    \text{ subject to } \sum_{j: e \in P_j} x_j \leq 1 \quad \forall e \in E. 
\end{equation}
\iffalse
\begin{equation} 
\begin{aligned}
& \max_{x} & & \sum_j g(x_j) \\
& \text{subject to}
& & \sum_{j: e \in P_j} x_j \leq 1 & \forall e \in E,\\
& & & x_j \in [0,b_j] & \forall j.\\
\end{aligned}
\end{equation}
\fi
In \cite{cao2022online}, a sufficient condition of having a roughly linear (in $|E|$) competitive algorithm for ONUM is presented. Together with our results, this implies the following corollary.

\begin{corollary}
Given any $\alpha$-competitive algorithm $\mathcal{O}$ for ONUM, Algorithm~\ref{alg:simple} using $\mathcal{O}$ as a subroutine implies a $(2,1)$-consistent, $2\alpha$-robust, and $3/2$-feasible learning-augmented online algorithm.
\end{corollary}

\subsection{Online Optimization under Inventory Constraints}

The online optimization with inventory constraints (OOIC) problem is introduced in \cite{lin2019competitive}. In this problem, a decision maker sells inventory across an interval of $T$ discrete time slots to maximize the total revenue. At time $t \in [T]$, the decision maker observes the revenue function $g_t: \nnreals \to \nnreals$ and makes an irrevocable decision on the quantity $x_t$. Upon choosing $x_t$, the decision maker receives revenue $g_t(x_t)$. The goal is to approximately maximize the total revenue, while respective the inventory constraint $\sum_{t \in T}x_t \le \Delta$ with a given parameter $\Delta > 0$. It is further assumed that for all $t \in [T]$, $g_t$ has the following nice properties: (1) $g_t$ is concave, increasing, and differentiable over $[0,\Delta]$, (2) $g_t(0) = 0$, and (3) $g_t'(0) > 0$ and $g'(0) \in [m, M]$ for some $M \ge m > 0$. OOIC has the following packing formulation:
\begin{equation}
    \text{maximize } \sum_{t \in [T]} g_t(x_t) 
    \text{ over } x_t \ge 0 \quad \forall t \in [T] 
    \text{ subject to } \sum_{t \in [T]} x_t \leq \Delta. 
\end{equation}
\iffalse
\begin{equation} 
\begin{aligned}
& \max_{x} & & \sum_{t \in [T]} g_t(x_t) \\
& \text{subject to}
& & \sum_{t \in [T]} x_t \leq \Delta,\\
& & & x_t \ge 0 & \forall t \in [T].\\
\end{aligned}
\end{equation}
\fi
The work \cite{lin2019competitive} presents a tight $\pi$-competitive algorithm for OOIC, where $\pi \in [\log(M/m)+1, \eta(\log(M/m)+1)]$ and $\eta:=\sup_{g,x \in [0,\Delta]} \{g'(0)x/g(x)\}$. Together with our results, this implies the following corollary.
\begin{corollary}
Given any $\alpha$-competitive algorithm $\mathcal{O}$ for OOIC, Algorithm~\ref{alg:simple} using $\mathcal{O}$ as a subroutine implies a $(2,1)$-consistent, $2\alpha$-robust, and $3/2$-feasible learning-augmented online algorithm.
\end{corollary}

\section{Empirical Evaluations} \label{sec:Experiments}
In this section, we present the applicability of our simple switching algorithm on real and synthetic data sets. Our focus will be on online packing LPs due to their simplicity which is sufficient to capture the key insights of our overall framework. 

\noindent\textbf{Datasets.}
For our synthetic data sets, we set the constraint matrix $A \in (\{0\} \cup [\ell,1])^{n \times n}$ and the upper bound matrix $b \in (0,1]^n$ with entries drawn independently and uniformly at random from $(0,1]$. Here, $\ell \in (0,1)$ is a lower bound parameter which ensures that $A$ is well-conditioned. Entries in $A$ are drawn uniformly at random from $[0,1]$. If $a_{ij} < \ell$, we round it to zero. We set $n=500$, $\ell = 0.01$, and $B=1$ to run our switching algorithm. All experiments are done in a 2018 HP Envy Laptop Model 17t-bw000 with 16 gigabytes of RAM and implemented in Matlab R2021a.

\noindent\textbf{Predictions.} We consider two types of predictions: (1) first, find the optimal offline solution $x$ by solving the full packing LP and then noisily corrupt the entries of $x$ by setting the entries to be $0$ independently with probability $p$. This is the same as the \emph{replacement rate} strategy in \cite{bamas2020primal,grigorescu2022learning}. (2) We consider related problem instances over time steps where the constraint matrix $A$ perturbs over time while $b$ is fixed. We have matrices $A_0, A_1, \ldots$ where $A_t$ is the constraint matrix at time $t$. $A_0$ is the starting synthetic data and $A_{t+1}$ updates $A_t$ by replacing $2n$ random entries with new entries. Three types of predictions are used: the \emph{batch prediction} which denotes the optimal solution when the constraint matrix is $A_0$, the \emph{online prediction} which denotes the optimal solution obtained from $A_{t-1}$ when the time is $t$, and the \emph{partial online prediction} which follows the online prediction by using the replacement rate strategy with $p=0.5$. A similar style of predictions, although not in an online context, has been employed in \cite{ChenEILNRSWWZ22, ErgunFSWZ22, DinitzILMV21, chen2022}.

\begin{figure}[h!]
\centering
\begin{subfigure}{0.49\textwidth}
    \includegraphics[width=\textwidth]{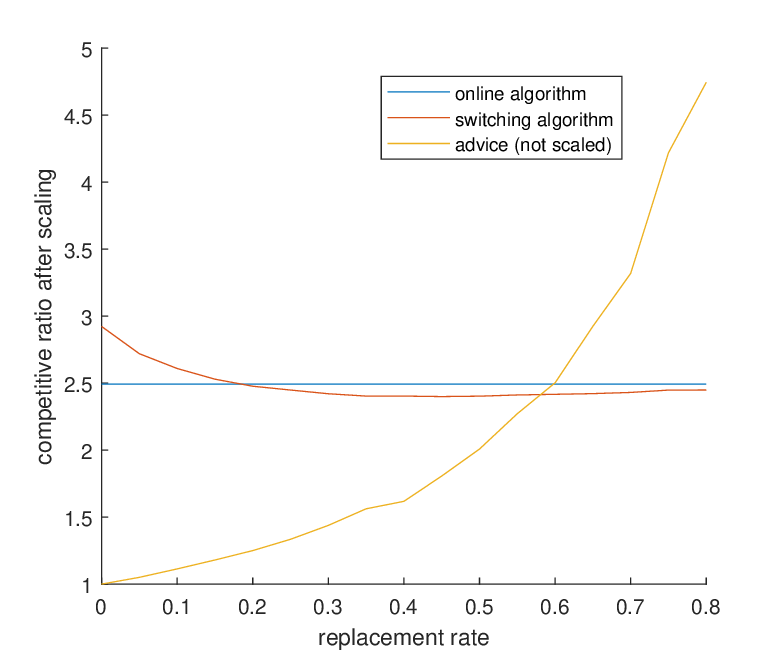}
    \caption{Competitive Ratio and Replacement Rate}
    \label{fig:first}
\end{subfigure}
\hfill
\begin{subfigure}{0.49\textwidth}
    \includegraphics[width=\textwidth]{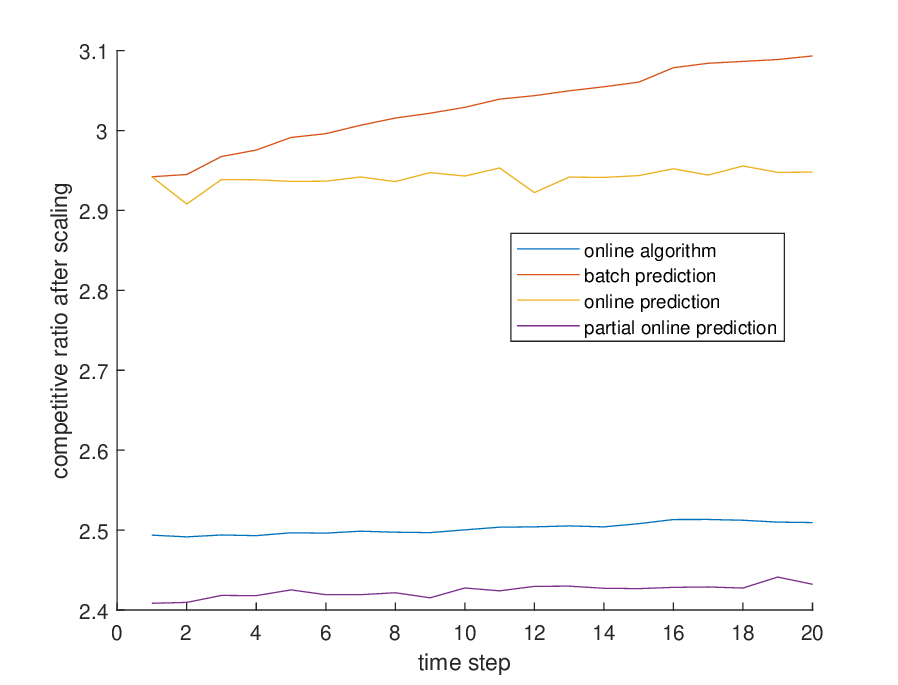}
    \caption{Competitive Ratio over Time Steps}
    \label{fig:second}
\end{subfigure}
\caption{Results on our synthetic data sets.}
\label{fig}
\end{figure}

\paragraph{Results.} 
Our results are shown in Figure \ref{fig}. Recall that our switching algorithm might generate an infeasible solution for packing LPs. We use the competitive ratio \emph{after scaling} to evaluate the performance of an online solution. Suppose the online solution $x$ is such that $Ax = b'$. We scale down $x$ by dividing $\max_{i=1}^n\{b'_i/b_i\}$ so that $x$ becomes feasible. We compute the competitive ratio w.r.t. the scaled online solution, which evaluates the \emph{variable distribution} performance of the online algorithm. Figure \ref{fig:first} shows the competitive ratios after scaling with different replacement rates $p$. Figure \ref{fig:second} shows the competitive ratios after scaling when the constraint matrix $A$ changes over time. For both settings, we consider 1000 online instances of the synthetic dataset and take the average of the competitive ratios after scaling.

In Figure \ref{fig:first}, the competitive ratio of the advice is not after scaling because, with a large corruption rate, $Ax'$ might have zero entries, where scaling is not desirable. The figure shows a smooth trade-off in the competitive ratio after scaling and the replacement rate. The switching algorithm outperforms the online algorithm and the advice when the replacement rate is high. With a low replacement rate, the switching algorithm heavily relies on the advice, resulting in a worse constant factor for the competitive ratio. The best type of advice is with a replacement rate of roughly 0.5.

In Figure \ref{fig:second}, the performance ranked from the worst to the best is as follows: the switching algorithm with the batch prediction, the switching algorithm with the online prediction, the pure online algorithm, and the switching algorithm with the partial online prediction. The partial online prediction advice intuitively is the best one based on the observation from Figure \ref{fig:first}.

\paragraph{Summary.} 
We summarize our experimental results: 
(a) Our theory is experimentally predictive and qualitatively validates the robustness and consistency of the switching algorithm.
(b) The switching algorithm is efficient to carry out in a black-box manner and execute in practice. 
(c) The switching algorithm can be applied to dynamic datasets varying over time.

\section{Further Discussion}\label{sec:discussion}

In this work, we demonstrated how a possibly folklore algorithm has wide applicability to the general problem of online algorithms augmented with advice, for packing problems with concave objectives. This work is motivated by the fact that a similar algorithm was initially missed when considering the dual problem to packing, namely covering problems, in the online
settings augmented by advice \cite{bamas2020primal, grigorescu2022learning}. Below we include some comparisons between the two dual online LP problems and approaches, which may help in a better understanding of the  space of problems solvable by algorithms that  use  classical online  algorithms as black-boxes. 

General online covering LPs take the following form:
\begin{align}
  \begin{aligned}
    \text{minimize } & c^T x
    \text{ over } x \in \nnreals^n 
    \text{ subject to } A x \geq \mathbf{1}.
  \end{aligned} \label{opt:covering}
\end{align}
where $A \in \nnreals^{m\times n}$ consists of $m$ \emph{covering} constraints, $\mathbf{1}$ is a vector of all ones, and $c \in \mathbb{R}_{>0}^n$ denotes the positive coefficients of the linear cost function. In the online setting, the cost vector $c$ is given offline, and each of the covering constraints, corresponding to rows of the constraint matrix $A$, is presented one-by-one online. The goal is to update the covering variables $x$ in a non-decreasing manner so that all constraints are satisfied on arrival, and the objective $c^T x$ is approximately minimized. A learning-augmented algorithm in this setting receives an advice $x' \in \nnreals^n$ on the suggested values of the covering variables. 

The simple switching algorithm in~\cite{grigorescu2022arxiv}(Appendix B) follows the online solution initially, and switches to the advice once the online cost surpasses that of the advice, taking the coordinate-wise maximum between the two candidate solutions to preserve the monotonicity of the covering variables. 
We remark that their algorithm ``switches'' between the advice and the online algorithm in a subtly different sense than ours.
\Cref{alg:simple} takes the average between the advice and the online solution initially, ``switching'' to using only the online solution when the advice becomes infeasible; \cite{grigorescu2022arxiv}'s learning-augmented algorithm follows strictly the online solution before ``switching'' to the advice once the online cost is higher.

Nonetheless,~\cite{grigorescu2022arxiv}'s setting and ours share several features. In particular, the switching strategy preserves consistency, robustness, and feasibility for both online covering and packing LPs. We hypothesize that features such as monotonicity, Lipschitz continuity, smoothness, convexity, or concavity of the objective, may be relevant to answering the main conceptual question raised by our work (see \cref{sec:intro}).

\paragraph{Acknowledgements} We are grateful to Kent Quanrud, Sandeep Silwal, Paul Valiant, and Samson Zhou for helpful discussions. 

\bibliographystyle{alpha}
\bibliography{arXiv_Main}

\end{document}